\documentclass[12pt,thmsa]{article}%
\usepackage{sw20elba}
\usepackage{amsmath}
\usepackage{graphicx}%
\usepackage{amsfonts}%
\usepackage{amssymb}

\begin{document}

\begin{center}
\textbf{Superconducting Phase Diagram of Li Metal in Nearly\linebreak
Hydrostatic Pressures up to 67 GPa}\bigskip

Shanti Deemyad and James S. Schilling

\bigskip

\textit{Department of Physics, Washington University\textbf{\linebreak }C. B.
1105, One Brookings Dr., St. Louis, MO 63130}

\bigskip
\end{center}

\noindent{\small Abstract. The dependence of the superconducting transition
temperature }$T_{c}${\small  on nearly hydrostatic pressure has been
determined to 67 GPa in an ac susceptibility measurement for a Li sample
embedded in helium pressure medium. \ With increasing pressure,
superconductivity appears at 5.47 K for 20.3 GPa, }$T_{c}${\small  rising
rapidly to }$\sim${\small  14 K at 30 GPa. \ The }$T_{c}(P)$%
{\small -dependence to 67 GPa differs significantly from that observed in
previous studies where no pressure medium was used. \ Evidence is given that
superconductivity in Li competes with symmetry breaking structural phase
transitions which occur near 20, 30, and 62 GPa. \ In the pressure range 20 -
30 GPa, }$T_{c}${\small  is found to decrease rapidly in a dc magnetic field,
the first evidence that Li is a type I superconductor.}\vspace{1cm}

Of the 52 elemental solids known to be superconducting, 23 enter this state
only if compressed under sufficiently high pressures \cite{ashcroft1}.\ The
latest confirmed member of this ever growing family is the alkali metal Li, an
element devoid of superconductivity at ambient pressure to temperatures as low
as 4 mK \cite{lang1}. Following an early report of possible pressure-induced
superconductivity in Li near 7 K \cite{lin1}, recent electrical resistivity
studies by Shimizu \textit{et al} \cite{shimizu1} to 48 GPa (480 kbar)
followed by ac susceptibility and resistivity studies by Struzhkin \textit{et
al} \cite{struzhkin1} to 80 GPa confirmed the onset of a superconducting state
for pressures above 20 GPa. The observed dependences of $T_{c}$ on pressure in
these three experiments, however, are in poor agreement. \ This may result
from the fact that no pressure medium was used, the ultrahard diamond anvils
\cite{shimizu1,struzhkin1} or boron-nitride spacers \cite{lin1} pressing
directly onto the Li sample, subjecting it to shear stress and plastic flow.
Shear-stress effects on $T_{c}(P)$ are well known from studies on such diverse
superconducting materials as organic metals \cite{organic1}, high-$T_{c}$
oxides \cite{schilling1}, MgB$_{2}$ \cite{schilling2}, and Re metal
\cite{chu1}; $\beta-$Hg, a shear-stress-induced body-centered tetragonal
modification of ordinary rhombohedral $\alpha-$Hg, exhibits distinctly
different superconducting properties \cite{schirber}. In a substance like Li,
where a multitude of potential phases lie very close in energy
\cite{neaton1,hanfland1}, the shear stress may be sufficient to induce
structural phase transitions.

Shimizu \textit{et al} \cite{shimizu1} reported that $T_{c}$ decreases only
slightly in moderate magnetic fields, a field of 30,000 Oe being required to
completely suppress Li's superconductivity at 34 GPa. \ This would indicate
that Li is a type II superconductor, an unexpected result for an elemental
superconductor, particularly for a simple (\textit{s,p}-electron) metal, like Li.

Neaton and Ashcroft \cite{neaton1} have obtained the counterintuitive result
that under sufficient compression the electronic structure of \textit{fcc} Li
departs radically from free-electron-like behavior. The anomalous increase in
the magnitude of the pseudopotential (and the electron-ion interaction) in the
\textit{fcc} phase under pressure leads not only to the possibility of a
superconducting state in Li, one where $T_{c}$ might increase significantly
with pressure, but also to possible structural transitions to phases with
reduced symmetry, such as a Li-ion pairing phase \cite{neaton1} or the
reduced-symmetry $hR$1 or $cI$16 phases observed by Hanfland \textit{et al}
\cite{hanfland1} for Li near 180 K.\ In Li, therefore, superconductivity and
symmetry breaking phase transitions are expected to compete with each other
\cite{neaton1} and possibly lead to an anomalous variation in $T_{c}$ with
pressure. Other superconducting properties, such as whether type I or type II,
values of the critical field(s), gap size, \textit{etc.}, may also be
anomalous and should be determined.

\textit{Ab initio} electronic structure calculations by Christensen and
Novikov \cite{novikov1} predict that under compression\textit{\ fcc} Li should
exhibit superconductivity where $T_{c}$ increases rapidly with pressure,
reaching values as high as 50-70 K.\ An increase in $T_{c}$ under pressure is
highly anomalous for a simple (\textit{s,p}-electron) metal, like Li. \ For
all known ambient-pressure simple-metal superconductors, $T_{c}$ is found to
\textit{decrease} under pressure because lattice stiffening effects normally
dominate over electronic effects \cite{schilling1}. The accurate determination
for Li of the intrinsic dependence of $T_{c}$ on pressure is, therefore, of
considerable importance.

In this paper we present an extensive determination of the superconducting
phase diagram, $T_{c}(P),$ to 67 GPa for a Li sample surrounded by helium, the
most hydrostatic pressure medium known. $T_{c}$ is derived from the magnetic
signature in the ac susceptibility at 1000 Hz and 3 Oe rms. Over the pressure
range to 67 GPa, the pressure dependence $T_{c}(P)$ differs significantly from
the results of earlier studies where no pressure medium was used
\cite{lin1,shimizu1,struzhkin1}. After the onset of superconductivity at 5.47
K for 20.3 GPa, $T_{c}$ rises rapidly to $\sim$ 14 K at 30 GPa, followed by an
abrupt change in the sign of $dT_{c}/dP$, signalling a likely phase transition
from \textit{fcc} to a lower symmetry structure. This transition at 30 GPa
appears to illustrate the competition between the superconducting state and
symmetry breaking phase transitions suggested above. In addition, $T_{c}$ is
found to decrease rapidly with applied dc magnetic field, giving the first
evidence that Li is a type I superconductor.

Because Li reacts readily with oxygen, water vapor, and nitrogen, all steps of
the sample preparation and pressure cell loading were carried out in an Ar-gas
glove box with continuous purification. High pressures were generated using a
diamond-anvil cell (DAC) made from nonmagnetic CuBe. Further experimental
details of the DAC and ac susceptibility measurement have been reported
previously \cite{schilling2}.

In the present experiment the miniature Li samples were cut from foil (Alfa
Aesar 99.9\%) and placed in a 250 $\mu$m dia. hole drilled through the center
of a preindented rhenium gasket (see Fig. 1). Tiny ruby spheres allow the
determination \cite{mao1} of the pressure \textit{in situ} to $\pm$ 0.2 GPa at
20 K. Before sealing the gasket hole shut with the opposing diamond anvils,
the hole is flooded with liquid helium which acts as pressure medium. The
pressure was changed at temperatures in the range 150 - 180 K; the Li sample
was kept at temperatures below 180 K during the entire high-pressure
experiment.\ After the conclusion of the experiment, the Li sample and rhenium
gasket were carefully examined. The preindented gasket thickness was found to
be reduced from the original 70 $\mu$m to 35 $\mu$m which is still greater
than that of the Li sample (20 - 25 $\mu$m). That neither the diamond anvils
nor the gasket wall pressed directly onto the Li sample can be seen in Fig. 1
(lower right) and can also be inferred from the irregular surface and
unaltered shape of the sample after the experiment.

In the present experiment, no superconducting transition was observed for Li
in the measured temperature region 4 - 60 K for applied pressures of 2, 11.5,
19.2, and 19.7 GPa.\ However, as seen in Fig. 2, at a pressure of 20.3 GPa a
clear superconducting transition appears at 5.47 K which increases rapidly
with pressure to $\sim$ 14 K at 30.2 GPa before falling and rising again above
50 GPa.\ The magnitude of the transition is consistent with 100\% shielding.
In Fig. 3 we summarize the results of our two extensive experiments
determining the superconducting phase diagram of Li metal. In the first
experiment (run 1) the pressure was first increased in 17 steps to 67 GPa
(solid circles) and then decreased\ in 13 steps to ambient pressure (open
circles), both in monotonic fashion. In run 2 (solid triangles) the pressure
was increased in 8 steps monotonically to 36.5 GPa; a blockage in the cooling
line prevented further experimentation. The reproducibility of the data for
runs 1 and 2 is seen to be excellent; the $T_{c}(P)$-dependence for increasing
or decreasing pressure shows only a small pressure hysteresis.

We now discuss in more detail the salient features exhibited by the
superconducting phase diagram in Fig. 3. The sudden appearance of
superconductivity in Li at 5.47 K for 20.3 GPa is likely due to a transition
from the low-temperature \textit{Rh}6 phase to the \textit{fcc} structure, as
suggested previously \cite{lin1,shimizu1,struzhkin1}. Increasing the pressure
above 20 GPa, $T_{c}$ is seen to rise rapidly at the rate $dT_{c}%
/dP\approx+0.9$ K/GPa, comparable to previous findings \cite{struzhkin1}, to a
value near 14 K at 30 GPa. Above this critical pressure the derivative
abruptly changes sign to $dT_{c}/dP\approx$ -0.4 K/GPa, giving clear evidence
for a second pressure-induced structural phase transition at 30 GPa.\ This
conclusion is reinforced by the reversible broadening of the superconducting
transition in the region 30 - 36 GPa. In view of the diffraction results on Li
at 180 K by Hanfland \textit{et al} \cite{hanfland1}, we tentatively identify
this second transition as \textit{fcc} to \textit{cI}16, although a transition
from\textit{\ fcc} to another phase, such as the intermediate \textit{hR}%
1\textit{\ }phase, is certainly possible. The data near 50 GPa is not
sufficiently dense to allow one to speculate whether the minimum in $T_{c}(P)$
near this pressure is characteristic for a single phase or arises from a
structural phase transition. The abrupt disappearance of superconductivity
above 62 GPa, however, signals a further phase transition, perhaps from
\textit{cI}16 to some unknown phase. Diffraction experiments on Li at
temperatures below 50 K under nearly hydrostatic pressure conditions are
clearly needed to arrive at an unequivocal structure assignment in the
low-temperature region.

Whatever the exact nature of the structural phase transitions indicated by the
data in Fig. 3 for pressures above 20 GPa, it is almost certain that they are
from the high symmetry \textit{fcc} structure to phases of lower symmetry,
such as \textit{cI}16 or \textit{hR}1. The highly anomalous increase in
$T_{c}$ with pressure between 20 and 30 GPa in the \textit{fcc} phase agrees
with the trend indicated in the calculation of Christensen and Novikov
\cite{novikov1} and bears witness to the large enhancement in the
pseudopotential and the electron-ion interaction with pressure predicted
earlier by Neaton and Ashcroft \cite{neaton1}. We speculate that with
increasing pressure the electron-ion interaction in Li is enhanced to the
extent that at 30 GPa a structural transition from \textit{fcc} to a lower
symmetry structure finally occurs. This transition removes, or at least
relieves, the anomalous enhancement of the pseudopotential ocurring in the
\textit{fcc} phase, allowing $T_{c}$ to decrease with pressure due to dominant
lattice stiffening effects, as in all canonical simple-metal superconductors
\cite{schilling1}. As the pressure is increased further, however, the story
repeats itself and $T_{c}$ passes through a minimum at 50 GPa and increases
again due to a renewed enhancement of the pseudopotential, only to fall again
above 62 GPa following a further symmetry lowering phase transition. To test
this scenario, detailed electronic structure calculations for the phases
actually present at low temperatures would be very useful.

The anomalous behavior of $T_{c}(P)$ for the ``simple'' \textit{s,p}-metal Li
arises from the close proximity of the electron cores of neighboring ions
under high compression. This situation will not be relieved, and Li allowed to
revert to its former free-electron-like behavior, before such astronomically
high pressures (far higher than those in the present experiment!) are applied
as to break up the atomic shell structure itself, resulting ultimately in a
Thomas-Fermi electron gas.

To further characterize the superconducting state of Li, in run 2 we subjected
the sample to dc magnetic fields up to 260 Oe.\ In the inset to Fig.
4,$\ \ T_{c}$ at 21.9 GPa pressure is seen to decrease from 5.95 K to 5.21 K
when a field of only 200 Oe is applied, yielding the initial critical field
slope $(dH_{c}/dT)_{T_{c}}\simeq$ -270 Oe/K. Since the applied fields are
insufficient to directly determine the critical field at 0 K, $H_{c}(0)$, from
the present data, we use the expression $H_{c}(0)=-%
\frac12
T_{c}(dH_{c}/dT)_{T_{c}},$ derived from the standard empirical relation
$H_{c}(T)=H_{c}(0)[1-(T/T_{c})^{2}]$ \cite{schmidt}, to obtain the estimate
from the 21.9 GPa data that $H_{c}(0)\approx$ -%
${\frac12}$%
(5.95 K)(-270 Oe/K) $=$ 800 Oe. Critical field data $H_{c}(T)$ at eight
further pressures were obtained, five of which are shown in Fig. 4. In the
pressure range 20 to 24 GPa the values of $(dH_{c}/dT)_{T_{c}}$, $H_{c}(0)$
and $T_{c}$ for Li are comparable to those found for canonical type I
simple-metal superconductors like Hg and Pb where $(dH_{c}/dT)_{T_{c}}\simeq$
-198.2 Oe/K and -237.3 Oe/K, $H_{c}(0)\approx$ 412 Oe and 803 Oe, and
$T_{c}\simeq$ 4.15 K and 7.195 K, respectively \cite{chanin,type1}. For $P<$
30 GPa, the evidence that Li is a type I superconductor is thus quite
compelling, although we cannot exclude the possibility of very weak type II
behavior. All known ambient-pressure simple-metal (\textit{s,p} electron)
superconductors are type I if in a pure and strain-free condition.

For pressures slightly above versus slightly below 30 GPa, it is interesting
to note that the initial slope $(dH_{c}/dT)_{T_{c}}$ increases approximately
twofold, likely signalling a comparable increase in $H_{c}(0)$. An abrupt
change in $(dH_{c}/dT)_{T_{c}}$ and/or $H_{c}(0)$ would be consistent with the
occurrence of a structural phase transition at 30 GPa. Since for type I
superconductivity we have the relation $H_{c}(0)\simeq%
\frac12
[\Delta(0)]^{2}N(E_{f})$ \cite{schmidt}$,$ such an increase in $H_{c}(0)$
could arise from an increase in the magnitude of either the superconducting
gap $\Delta(0)$ or the density of states $N(E_{f})$. Alternatively, the phase
transition at 30 GPa may create lattice defects which lead to weak type II
behavior and an increase in the (upper) critical field.

Shimizu \textit{et al} \cite{shimizu1} reported from resistivity measurements
at 34 GPa, where $T_{c}\simeq$ 7 K, that $H_{c}(0)\approx$ 30,000 Oe, a value
6$\times$ larger than the maximum value of $H_{c}(0)$ estimated in the present
experiment. This high value of $H_{c}(0)$ clearly points to type II
superconductivity for their Li sample. Since in their experiment the direct
contact of the Li sample with the diamond anvils and stiff gasket generated
relatively large shear stresses, it is conceivable that the resulting plastic
deformation led to a high density of defects with a sharp reduction in the
electronic mean-free-path, thus promoting type II behavior.

\vspace{1cm}

\noindent\textbf{Acknowledgments. \ }The authors thank J. Hamlin, A.-K. Klehe,
V.V. Struzhkin, V. Tissen, and especially S. Solin for critical reading the
original manuscript. \ Experimental support by J. Hamlin is acknowledged.
\ One of the authors (JSS) is grateful to N.W. Ashcroft and J.B. Neaton for
stimulating discussions. \ Research is supported by NSF grant DMR-0101809.

\noindent{\Huge Figure Captions}

\bigskip\ 

\noindent\textbf{Fig. 1. \ }(upper) Reflected-light photograph into Ar-gas
glove box of Au-plated rhenium gasket preindented with diamond anvil (0.5 mm
dia. culet). \ In 250 $\mu$m dia. hole are seen for run 2 at ambient pressure
the Li sample at 6 o'clock plus clusters of ruby spheres. (lower)
Transmitted-light photograph of hole containing Li sample at (left) ambient
pressure and (right) 30 GPa. All three photographs are to same relative scale.\bigskip

\noindent\textbf{Fig. 2. \ }Superconducting transition of Li in the real part
of the ac susceptibility (1000 Hz, 3 Oe rms) for 8 values of the
pressure.\ All data are to same scale but are shifted vertically for clarity.
All data are from run 1 for increasing pressure, except data at 20.3 GPa (run
2) and 43.0 GPa (run 1, decreasing pressure).\bigskip

\noindent\textbf{Fig. 3. \ }Superconducting phase diagram of Li metal under
nearly hydrostatic pressure: \ run 1 increasing pressure ($\bullet$), run 1
decreasing pressure ($\circ$), run 2 increasing pressure ($\blacktriangle$).
The error in pressure is $\pm$ 0.2 GPa. $T_{c}$ is determined by
superconducting midpoint to $\pm$ 50 mK; vertical ``error bars'' give
temperatures of the superconducting onset and completion. No superconducting
transition is observed above 4 K for pressures below 20 GPa or at 67 GPa.
Dashed lines are guides to eye with slopes $dT_{c}/dP\simeq$ +0.9 K/GPa (left)
and -0.4 K/GPa (right).\bigskip

\noindent\textbf{Fig. 4. \ }Data points ($\bullet,\bigstar$) give to 220 Oe
the critical magnetic field $H_{c}$ versus temperature at various pressures
for superconducting Li.\ Solid, dashed, and dotted lines are obtained using
the empirical expression $H_{c}(T)=H_{c}(0)[1-(T/T_{c})^{2}]$ to allow fits to
the present data for Li in comparison to ambient-pressure data for Pb. For
$P=$ 28.6, 31.8, and 36.5 GPa we find $H_{c}(0)\approx$ 2400, 5000, and 4500
Oe, respectively. Inset shows ac susceptibility data at 21.9 GPa for applied
magnetic fields of 0 and 200 Oe.
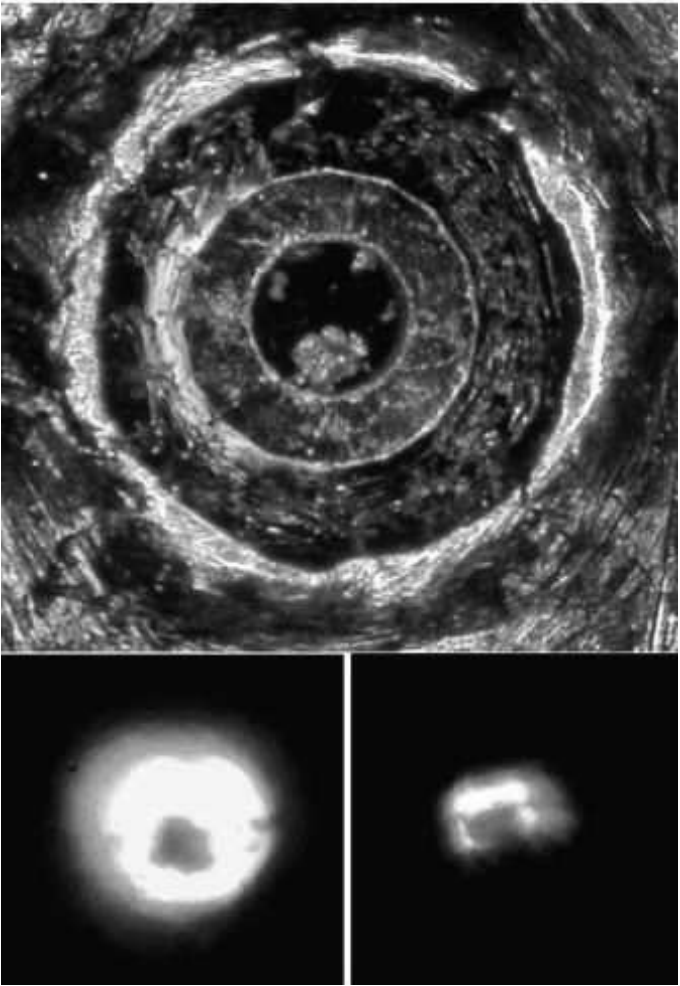
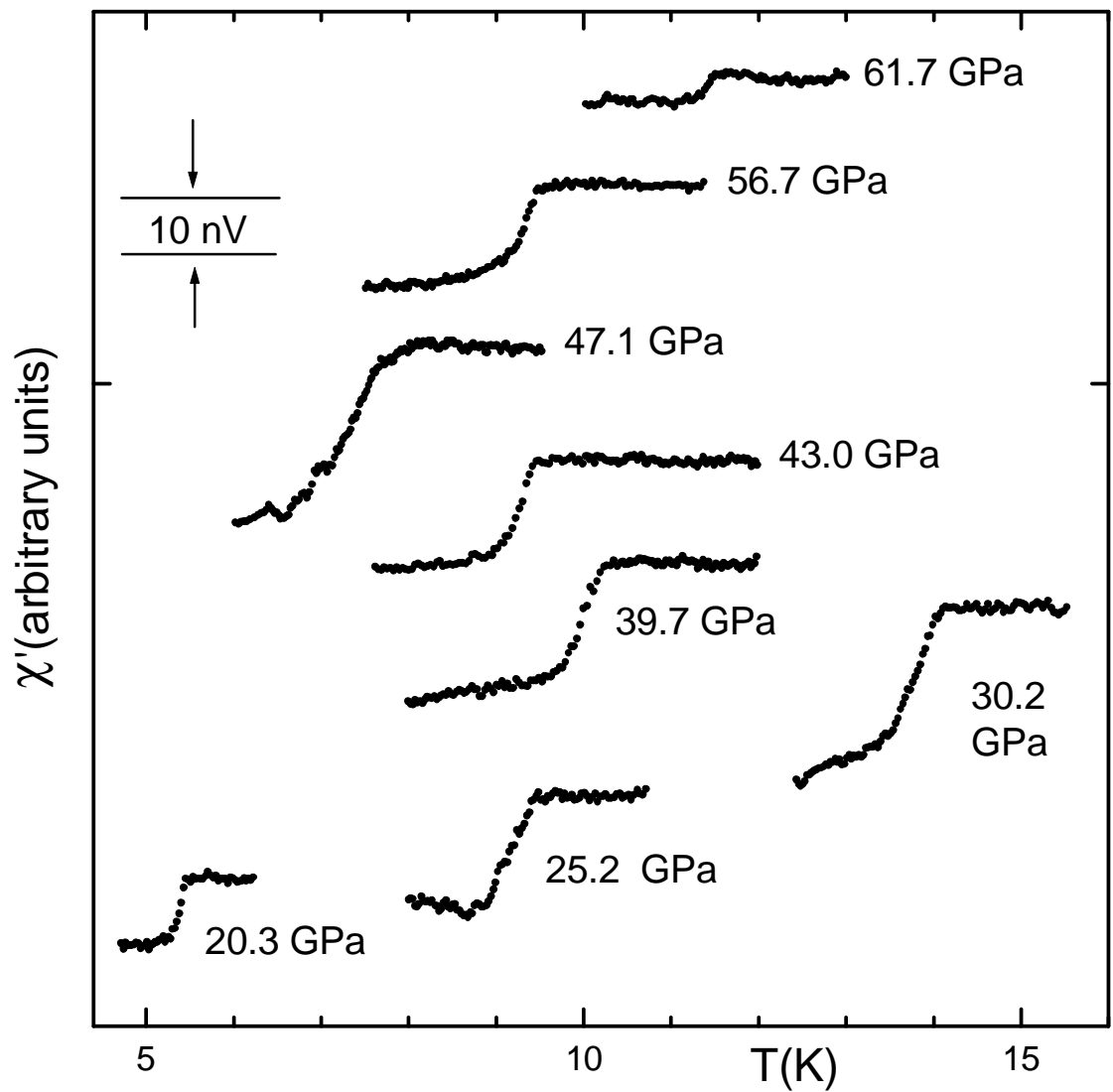
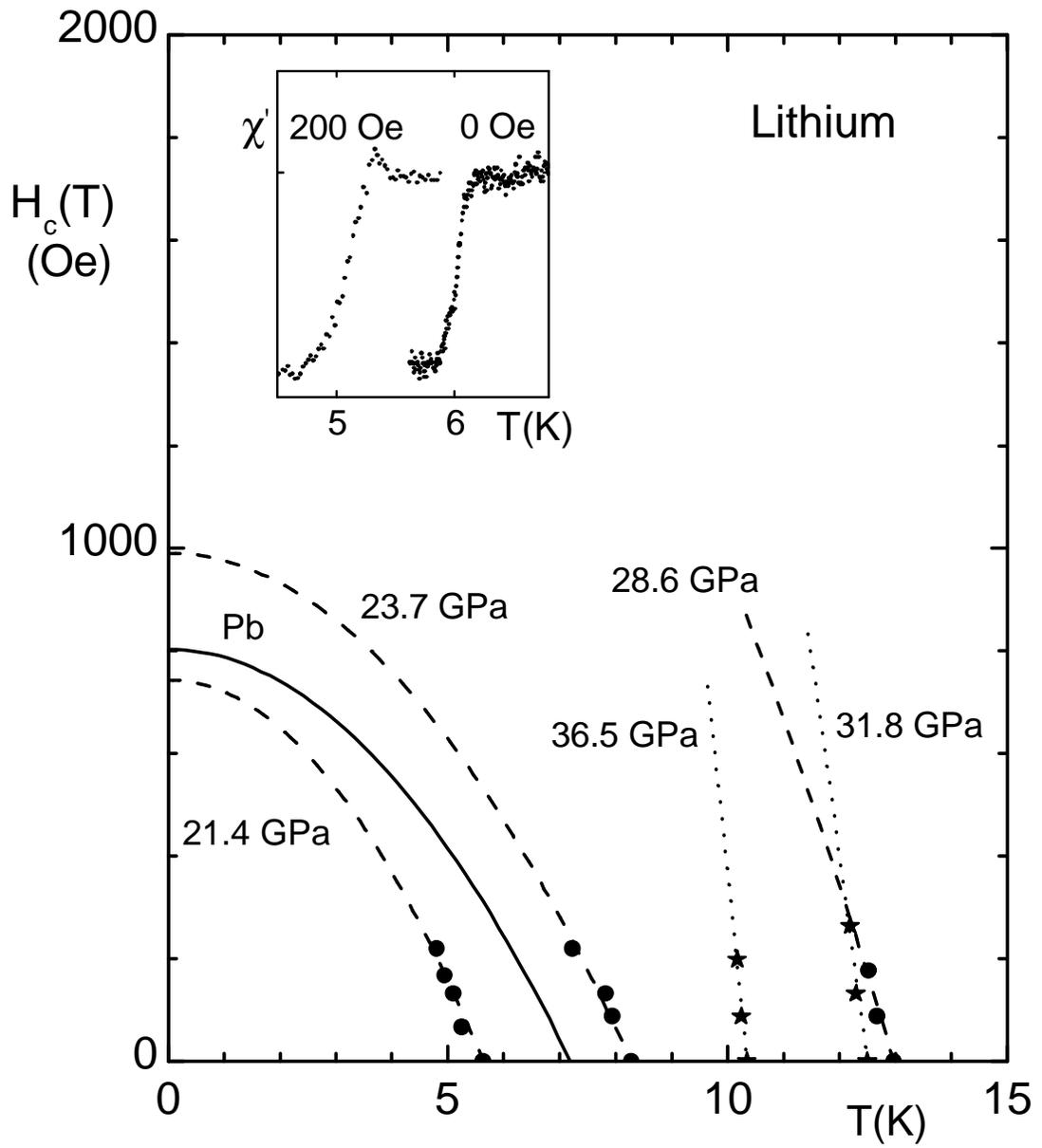
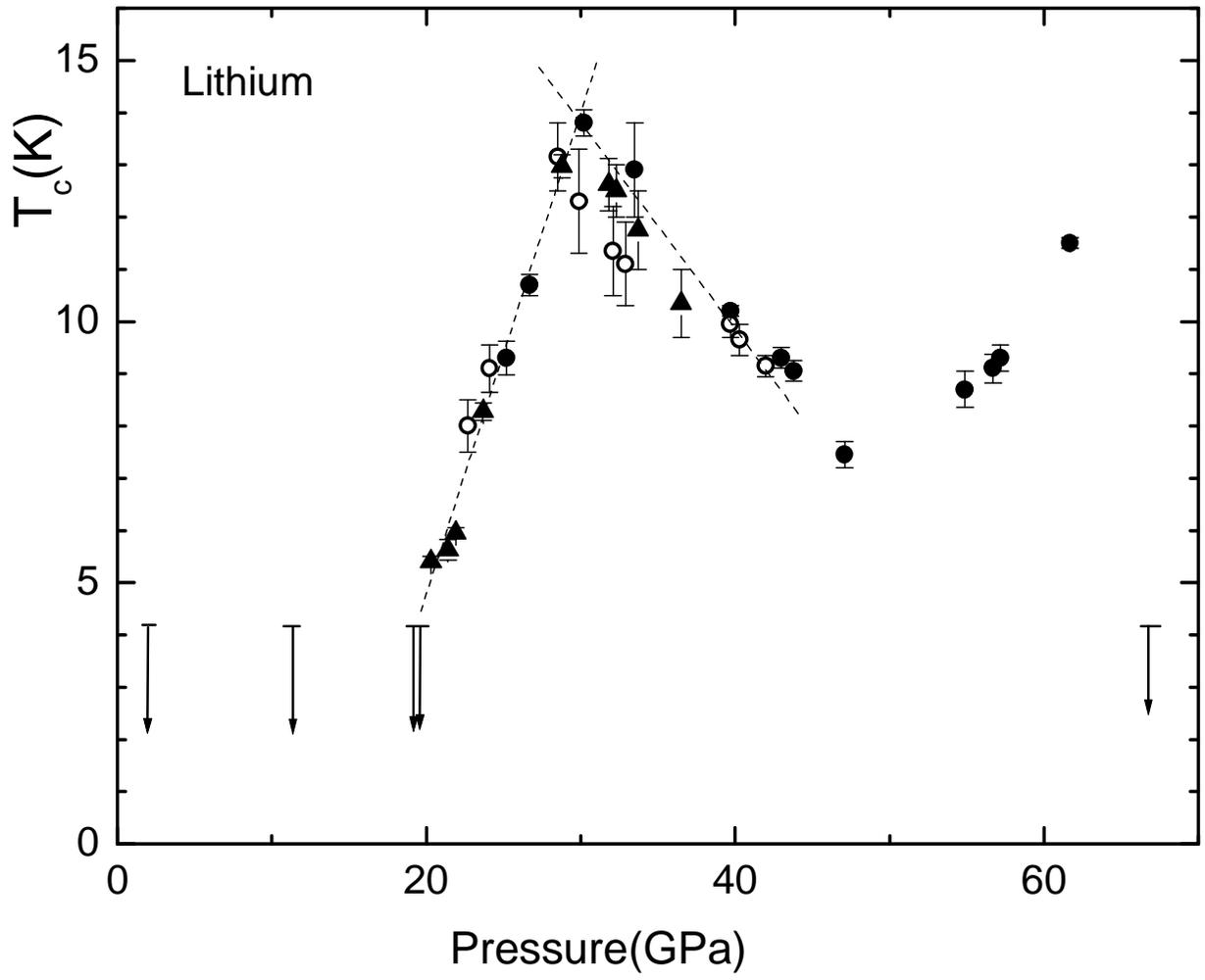
\end{document}